\pdfoutput=1

\documentclass{article}       

\usepackage[usenames, dvipsnames]{xcolor}
\usepackage{graphicx}
\usepackage{array}
\usepackage{multirow}
\usepackage{enumitem}
\usepackage[colorlinks=true, citecolor=blue, urlcolor=blue, linkcolor=blue]{hyperref}
\usepackage{amsmath}
\usepackage{caption}

\begin{document}

\title{Improvements on Fresnel arrays for high contrast imaging}

\author{{Roux} Wilhem \and {Koechlin} Laurent}


\date{Submitted: April 9, 2017 -- Accepted: December 17, 2017\\
by Experimental Astronomy}

\maketitle

\begin{abstract}
	
The Fresnel Diffractive Array Imager (FDAI) is based on a new optical concept for space telescopes, developed at Institut de Recherche en Astrophysique et Plan\'etologie (IRAP), Toulouse, France. For the visible and near-infrared it has already proven its performances in resolution and dynamic range. We propose it now for astrophysical applications in the ultraviolet with apertures from 6 to 30 meters, aimed at imaging in UV faint astrophysical sources close to bright ones, as well as other applications requiring high dynamic range. 
Of course the project needs first a probatory mission at small aperture to validate the concept in space. 
In collaboration with institutes in Spain and Russia, we will propose to board a small prototype of Fresnel imager  on the International Space Station (ISS), with a program combining technical tests and astrophysical targets. The spectral domain should contain the Lyman-$\alpha$ line ($\lambda= 121$ nm).
As part of its preparation, we improve the Fresnel array design for a better Point Spread Function in UV, presently on a small laboratory prototype working at 260 nm. Moreover, we plan to validate a new optical design and chromatic correction adapted to UV. 
In this article we present the results of numerical propagations showing the improvement in dynamic range obtained by combining and adapting three methods : central obturation, optimization of the bars mesh holding the Fresnel rings, and orthogonal apodization. 
We briefly present the proposed astrophysical program of a probatory mission with such UV optics.

\textbf{Keywords}: Fresnel arrays -- Diffractive optics -- UV imaging -- Apodization -- High dynamic range -- High angular resolution -- Exoplanets
	
\end{abstract}

\section{Introduction}

\begin{figure}[hbp]
	\centering
	\includegraphics{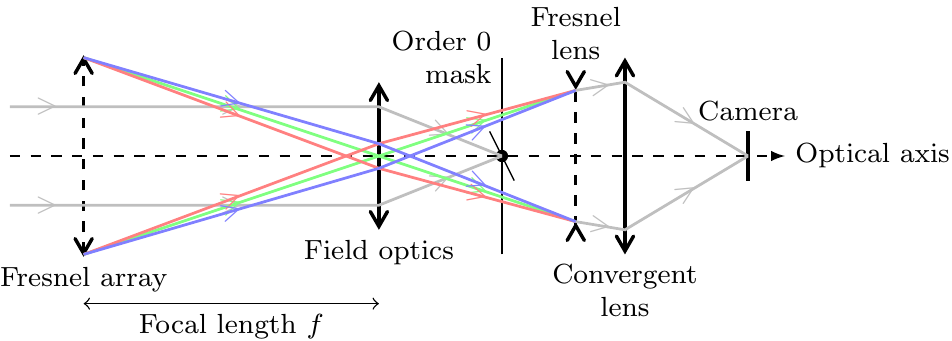}
	\caption {Scheme of the previous Fresnel imager prototype, adapted to the visible. The achromatic rays are in \textit{gray}. The diffracted rays in color represent different wavelengths to show chromatism: \textit{green} for the central wavelength in the bandwidth, \textit{red} for longer and \textit{blue} for shorter. This configuration is optimized for the central wavelength \textit{green}, as the field optics is placed at the corresponding focal plane. One can see that  order zero (plane wave) is blocked by the mask in the focal plane of the field optics. Credits: Paul Deba.}
	\label{fig:fresnel_imager}
\end{figure}

Focusing light by diffraction using an alternation of opaque and transparent rings ({\it i.e.} binary transmission) was first achieved by Soret (1875) \cite{Soret1875}, inspired by Fresnel experiments (1818) \cite{Fresnel1818}. From this were made Fresnel Zone Plates (FZP). The Fresnel Diffractive Array Imager (FDAI) uses diffractive optics as primary aperture: a Fresnel array, designed initially by Koechlin et al. (2005) \cite{Koechlin2005}. It is an adaptation of the FZP and uses a binary mask in which transmissive regions are just void apertures. Each opaque ring associated to its adjacent void ring corresponds to a Fresnel zone. $N$ -- number of zones along the radius -- sets its focal length $f$ that can be closely approximated by:
%
\begin{equation*}
	f \simeq \frac{{d_{array}}^2}{8 \lambda N}
\end{equation*}
where $d_{array}$ is the dimension (aperture side) of a square Fresnel array, and $\lambda$ the wavelength. So a FZP has chromatism: for broad band observation it requires correcting optics as proposed by Schupmann (1899) \cite{Schupmann1899}, and later by Faklis and Morris (1989) \cite{Faklis1989}. In our optical setup (see Figure \ref{fig:fresnel_imager}), this correction is made by secondary diffractive optics in a pupil plane, conjugate of the FZP causing aberration.

Fresnel arrays are very lightweight compared to classical optics, that's one of the reasons why they can be convenient in space. A very large aperture is feasible: up to 30 meters, (Hinglais 2011) \cite{Hinglais2011}, which could provide a higher angular resolution than present space telescopes. At small aperture the concept has been tested from the ground on astrophysical sources: high contrast binary objects (Sirius AB), Mars and its satellites, extended objects: the Moon, planetary surfaces, M42 nebula (Koechlin et al. 2011) \cite{Koechlin2011}. As light is focused only by diffraction through void apertures, it has no interaction with optical material like lenses or reflective surfaces. This is an advantage for UV, more absorbed and diffused by classical optics. Raksasataya et al. (2010) \cite{Raksasataya2010} made simulations showing examples in UV astrophysics, such as protoplanetary discs dominated by far-UV radiations. In 2010, a mission based on FDAI has been proposed to the European Space Agency (ESA) in the frame of ``Cosmic Vision", but was rejected, among other reasons, because its Technology Readiness Level (TRL) was not high enough. Since then, FDAI has progressed and should be ready for tests in relevant conditions: in space. We prepare a proposal for a mission on the International Space Station (ISS) (Roux and Koechlin 2015) \cite{Roux2015}. In addition to the technical tests, a scientific program will be defined in collaboration with the Universidad Complutense de Madrid (UCM) and the Institute of Astronomy of the Russian Academy of Sciences (INASAN).

In order to avoid diffused light, a Fresnel array has no optical material support, so it is a binary mask: thin opaque plate with void zones. The opaque zones are the union of Fresnel rings and orthogonal bars used to hold them in position. This whole pattern is contained in a square frame. The bar mesh is necessary but has a negative impact on the PSF. In addition to high angular resolution, the astrophysicists community needs instruments with high dynamic range capabilities to detect faint sources in the vicinity of the brightest objects of our Universe --- local or far --- for investigating new domains of science. This requirement have motivated us to improve the PSF of Fresnel arrays, before preparing a proposal for the probatory mission. 

Having no known analytic solution for such an intricate pattern as Fresnel rings with holding bars, in this article we focus on the optimization of Fresnel array design by numerical simulations of their PSFs (Point Spread Functions): their diffraction pattern in the focal plane. First, we describe the configuration of our laboratory prototype, the method we used for numerical simulations and the criteria for comparing performances. Then we study separately the optical improvements implemented and optimized on the Fresnel arrays of our laboratory prototype: central obturation, holding bars mesh and apodization. These simulations will of course not waive the requirement for optical tests in the future, but they greatly accelerate the optimization for our laboratory prototype. We will adapt and apply these solutions to the configuration of the future probatory mission on the ISS.

\section{Methods}

\subsection{Numerical simulations}
We use numerical light propagation to compute the PSFs of our Fresnel arrays before manufacturing them: we have no analytic solution, except in the case of relatively simple FZPs. Our codes to test Fresnel arrays have been initially developed by Serre (2010) \cite{Serre2010}, and we now have developed a software better suited to our new designs ({\it e.g.} regularly spaced bars, square apodization).

\begin{figure}[htbp]
	\centering
	\includegraphics{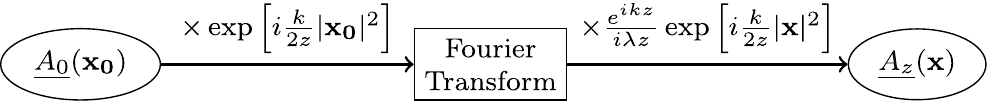}
	\caption{Summarized scheme of numerical propagation using Fresnel diffraction.}
	\label{fig:Fresnel_method}
\end{figure}

We use Fresnel propagation to compute the diffraction pattern at a finite distance $z$ from the initial (objective) plane. After simplifications, the complex amplitude of the light wave can be expressed by:
\begin{multline}
\underline{A_z} ({\bf x})=
\frac {e^{ikz}}{ i\lambda z } \exp { \left[ { i \frac { k }{ 2z } |{ \bf x}|^2}      \right]} \times
\\
\iint _{ -\infty}^{ \infty}{ \underline{A_0}({\bf x}_0)   
        \exp  \left[   i \frac { k }{ 2z } |{ \bf x_0}|^2     \right] 
		\exp  \left[  -i \frac { k }{ z } (\bf x \cdot x_0)  \right] }   d{\bf x}_0
\end{multline}
where $k=\frac{2\pi}{\lambda}$ is the module of the wave vector, $\bf x$ is the position vector in the target plane orthogonal to the optical axis, $\underline {A_z} (\bf x)$ is the complex amplitude in this plane, placed after propagation at distance $z$, and $\underline{A_0}(\bf{x_0})$ the amplitude in the aperture (objective) plane. The method is summarized in Figure \ref{fig:Fresnel_method}.

With this method we can compute PSFs rapidly compared to the direct application of Kirchhoff's diffraction formula, thanks to the focal length that is in agreement with the Fresnel approximations. Moreover, we take care that Fresnel arrays are sufficiently sampled to avoid undersampling on the narrowest rings.

To model the propagation through the whole Fresnel imager, we must compute the propagated wavefront plane by plane, \textit{i.e.} compute the complex amplitude of the wave on each optical element and apply the associated phase or mask term. According to Figure \ref{fig:fresnel_imager}, computing the PSF of the whole optical train with the old prototype involved five Fresnel propagations, but now only three are required: Fresnel array $\rightarrow$ field optics  $\rightarrow$  convergent Fresnel grating $\rightarrow$  focal plane. Indeed in our new UV setup we have been able to remove two elements, which are no longer necessary: the order zero mask and a converging lens, while keeping same or better performances. Furthermore, as the present study just compares among different designs of primary Fresnel arrays having the same optical layout downstream, modeling the first propagation only is sufficient: Fresnel array $\rightarrow$  prime focus. 



\subsection{Criteria for PSF comparison}

In our case, the PSF is the image of a point source at infinity on axis after a propagation through a Fresnel array to its focal plane of diffraction order 1. The simulated PSFs are images made of pixels associated to the relative irradiances. 


The orthogonal coordinates are expressed in resels. One resel is a length unit in our images, which corresponds to the main lobe half width (first zero of the diffraction pattern). 
As our Fresnel arrays have square apertures, a resel has a linear dimension of $\frac{\lambda f}{d_{array}}$ in the focal plane , or angularly $\frac{\lambda}{d_{array}}$.

\begin{figure}[htbp]
	\centering
	\begin{tabular}{cc}
		\includegraphics[width = 0.45\textwidth]{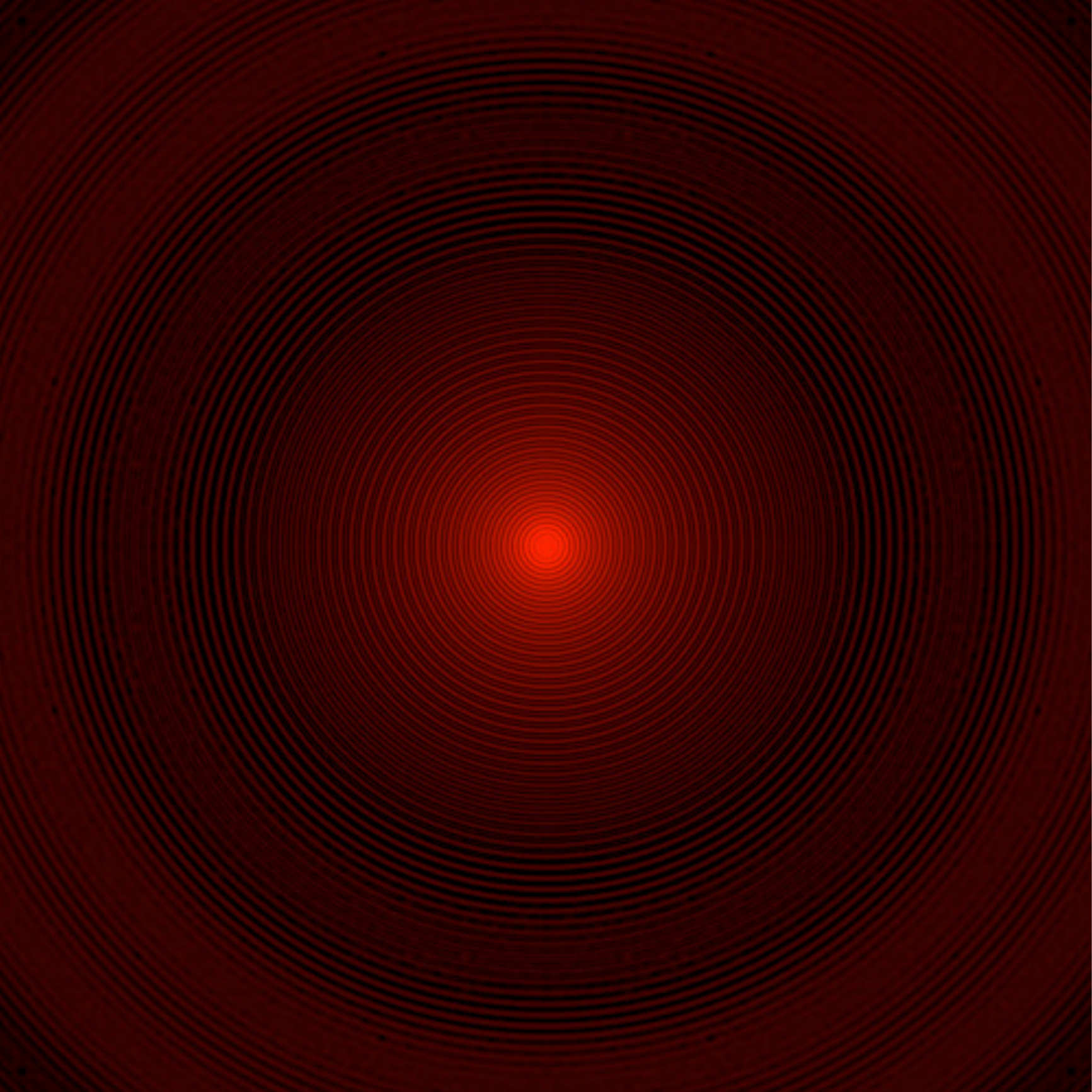} &
		\includegraphics[width = 0.45\textwidth]{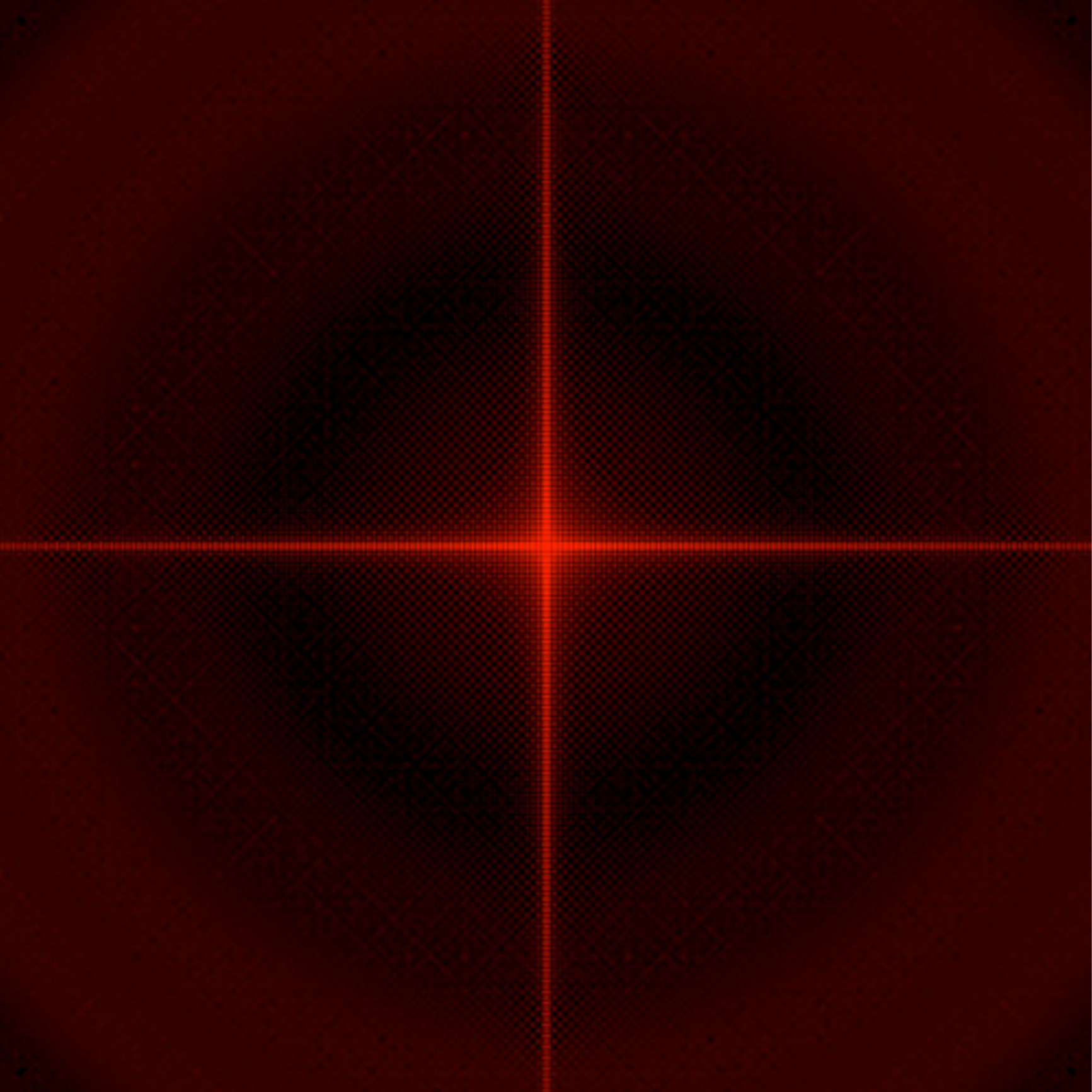} \\
	\end{tabular}
	\caption{Typical PSF at order 1 of a binary Fresnel zone plate, without bars, without apodization, and without central obturation. The display here boosts low luminosity regions, using a $\log$ lookup table. \textit{Left: }Circular aperture. The central lobe is surrounded by concentric rings. \textit{Right: }Square aperture. We can distinguish the central lobe, four orthogonal spikes due to the aperture edges, delimiting four cleaner fields. }
	\label{fig:typical_psf}
\end{figure}

In the PSF of a square Fresnel array we distinguish three different parts, as in Figure~\ref{fig:typical_psf} right: 
\begin{itemize}
\item a square central lobe: light focused by diffraction order 1, with a dimension of $2\times2$ resels;
\item four orthogonal spikes: due to the square aperture edges, delimiting four `clean field' quadrants;
\item four quadrants: containing (or not, depending on central obturation) stray light from the other diffraction orders, mainly order 0.
\end{itemize}
Thanks to this symmetry, we need only analyze one quadrant in each case to characterize the PSF quality. To assess the quality of images provided by Fresnel arrays, we use two criteria: transmission efficiency and dynamic range. Other criteria such as angular resolution are of course primordial but related to the size and wavelength, at a smaller extent to apodization (Myers, 1951) \cite{Myers1951}. They are less impacted by the internal layout.

\subsubsection{Efficiency of light transmission to the image}

We measure the efficiency of Fresnel arrays by the integrated light in the central lobe of their PSF 
\textit{i.e.} we normalize the integrated irradiance inside the central lobe, by the the integrated irradiance incident over the whole aperture: the blocking effect of the opaque zones is taken into account here.

This transmission efficiency is crucial: it drives the acquisition time required for a given S/N ratio. 
For a binary FZP it reaches asymptotically 10.1\% at maximum, for an infinite number of Fresnel zones $N$ in a finite aperture, no central obturation, no bars and no apodization (see a simple demonstration in Attwood (2007) \cite[Chapter~9]{Attwood2007}). 


\subsubsection{Dynamic range}

The dynamic range in the images drives the ability to detect faint sources next to bright ones. It is normally expressed by the maximum over the minimum value of a quantity. In our case
 it is computed from a PSF as the ratio between the maximum in the central lobe, and the highest `spurious' peak outside that lobe, in the clean field. 
 
In the case of an exoplanet orbiting around its star, we infer an angular distance range, so we define a region of interest in the clean field, outside the spikes. For circular apertures,
 regions of interest are usually defined as rings limited by two radii in resels: IWA (Inner Working Angle) and OWA (Outer Working Angle).
In our case with square apertures we define regions of interest by these same values, but they correspond to four squares  limited by $|x|,|y| \in [\mbox{IWA,OWA}]$, thus excluding the spikes. Due to the spikes, a single exposure lacks two narrow orthogonal bands in the field, but one can complete the coverage with another exposure rotated 45$^\circ$ (Koechlin, Serre, Duchon 2005) \cite{Koechlin2005}.

 Our measure is adapted to the situation of faint sources in the vicinity of bright, quasi punctual sources.
 In a different case, as an extended bright source covering the whole field (\textit{e.g.} dense stellar field or large planetary surface), 
 the dynamic range will be affected by the convolution by all the PSF's secondary peaks outside the central lobe, including the spikes. Experience shows (Koechlin et al. 2011) \cite{Koechlin2011}  that the images remain at high resolution, but no longer at high contrast. 

It's difficult to define  a region of interest a priori, because it depends on 
the target object and science case.
Thus, in this study we compute the dynamic range for different Fresnel arrays over the PSF up to 100 resels, 
and we also compute it in certain regions with more details. 

\subsubsection{A compromise between efficiency and dynamic range}

In the following we show that different designs of arrays at same dimensions improve the dynamic range but degrade the transmission (\textit{i.e.} luminosity), so we have to find a trade-off between these two. In astrophysical applications, their relative importance depends on the science case.

For example, direct imaging of exoplanets requires high dynamic range 
and that leads to less luminosity, thus requiring longer acquisition times or larger apertures. 
As studied by CNES (Hinglais 2011)\cite{Hinglais2011}, very large (30 m) Fresnel arrays could be envisioned as lightweight and foldable for launch, but the best compromise needs to be found for a given, affordable aperture.



In the case of our proposal for a probatory mission on the ISS (see Section \ref{s:iss}), the constraint will be the focal length: the distance between the Fresnel array and the field optics will be limited to $50$ m, which is the available length along the integrated truss (Roux and Koechlin 2015) \cite{Roux2015}. That maximum focal length, the minimum observed wavelength and the maximum number of Fresnel rings that can be engraved, will constrain the Fresnel array dimension. We expect it will be limited to 15 - 20 cm. As this validation mission is intended to have some, though modest, scientific return we have to carefully choose the balance between efficiency and dynamic range. Further, larger (if financed) missions will have the same constraints,  at a  larger scale.

In this study, we show the improvements in dynamic range, along with the associated efficiency effect on light transmission, on the laboratory prototype.

\subsection{Specifications of the laboratory prototype}

For this numerical study we take the specifications described in Table~\ref{table:Fresnel_array_specs}, corresponding to the Fresnel imager prototype built,  
which is our Mark III: `UV, lab'. The previous ones were respectively Mark I: `visible, lab' and Mark II: `visible, sky'. The specifications cannot be exactly the same as in a future `UV, sky' prototype because under $242$ nm the UV radiations are strongly absorbed by the O\textsubscript{2} in the air. Our ground prototype tests the UV optics at $260$ nm, as a long vacuum line housing the whole optical path is unfortunately not in our budget. 
As part of the new design it is necessary at first  to validate the new chromatic corrector for UV. A single reflective surface: a concave blazed Fresnel mirror, replaces the four surfaces and two transmissive media used previously in the Mark II prototype for that purpose, and for image formation at final focus. The new setup also departs from the previous ones by the use of a central obturation to suppress the unwanted orders of diffraction.

\begin{table}[htbp]
	\begin{center}
		\caption{Specifications of the test UV Fresnel array. We have chosen a $260$ nm wavelength to allow air propagation.}
		\label{table:Fresnel_array_specs}
		\begin{tabular}{|l|c|}
			\hline
			Studied wavelength $\lambda$ & 260 nm (Middle-UV)\\
			\hline
			Aperture size $d_{array}$ & 65 mm \\
			\hline
			Number of Fresnel zones (rings) $N$ & 160\\
			\hline
			Resulting focal length $f$ & 12.755 m \\
			\hline
		\end{tabular}
	\end{center}
\end{table}

\section{Central obturation}

\begin{figure}[htbp]
	\centering
	\includegraphics{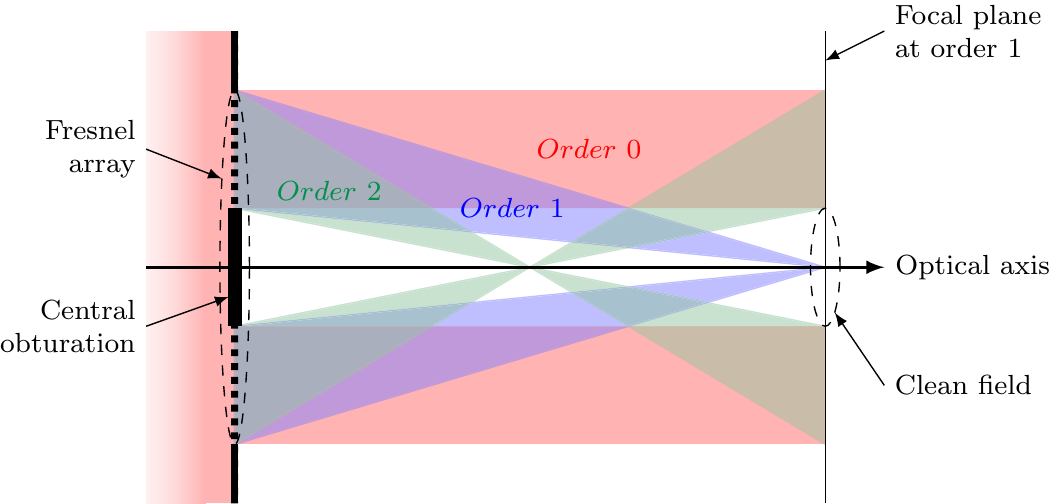}
	\caption[Sketch of consequences of a central obturation.]{ Geometrical effect of central obturation on background light: the focus at diffraction order 1 is in the shadow of the obturation for all other diffraction orders. Light from negative orders is divergent and light from positive orders above 1 is rejected far from the center of the PSF.}
	\label{fig:central_obt}
\end{figure}

The role of the order 0 mask in our previous setups was to keep the unfocused light (mostly the order 0 of diffraction) far away from the central lobe of the PSF, to get a clean field with high dynamic range. This mask was in the focal plane of the field optics, upstream form the pupil plane (Figure~\ref{fig:fresnel_imager}), and it was held by a thin spider. It blocked order 0 but induced stray light by diffraction around it and its spider. Our new prototype has a central obturation embedded in the primary array, which as before blocks order 0, and in addition all the unwanted orders of diffraction (different from 1), as shown in  Figure~\ref{fig:central_obt}.

A central obturation in the Fresnel array reduces stray light, simplifies the design, but also reduces inevitably light transmission. It has an altering effect on the PSF but we can reduce it by apodization, as tested on their design by Vanderbei et al. (2003) \cite{Vanderbei2003}. We could also place this obturation in a pupil plane -- atop the chromatic corrector -- but this requires a highly absorbing material to avoid reflections.

For extended objects, the image plane is no longer protected by the central obturation and some unfocused light from order zero reaches the image. However, large field imaging is still done at high (nominal) resolution and acceptable contrast, as shown by our tests on the moon surface (Koechlin {\it et al.} 2011).

\subsection{Shape and dimension of the central obturation}

Our central obturation
 is a square mask, which confines its diffraction into four spikes superposed to those from the square aperture.

The clean field, in the central obturation's shadow, has an angular extent conditioned by that obturation. Equations~\ref{eqn:mask_0_size} shows that this extent in resels has a simple expression involving: number of Fresnel zones $N$ and size ratio $\epsilon$.

\begin{equation}
	\begin{split}
		\epsilon = \frac{d_{obt}}{d_{array}} &= \frac{n_{resels}\times d_{resel}}{d_{array}} 
		\\
		& = \frac{n_{resels}}{d_{array}} \times \frac{\lambda f}{d_{array}}
		\\
		& = \frac{n_{resels}}{d_{array}} \times \frac{\lambda}{d_{array}} \times \frac{{d_{array}}^2}{8 \lambda N}
		\\
		& = \frac{n_{resels}}{8 N}
	\end{split}
	\label{eqn:mask_0_size}
\end{equation}

The parameters are:
\begin{itemize}
    \item $\epsilon$: ratio: central obturation to total aperture (linear);
	\item $d_{obt}$: size of the central obturation;
	\item $n_{resels}$: clean field expressed in resels;
	\item $d_{resel}$: resel size.
\end{itemize}

In our setup we choose  $n_{resels} = 200$, {\it i.e.} the a clean field extends 100 resels from center. This corresponds to a central obturation $d_{obt} = 10.2$ mm, obturation ratio $\epsilon = 0.156$.

\subsection{Results and discussion}

\begin{figure}[htbp]
	\centering
	\includegraphics{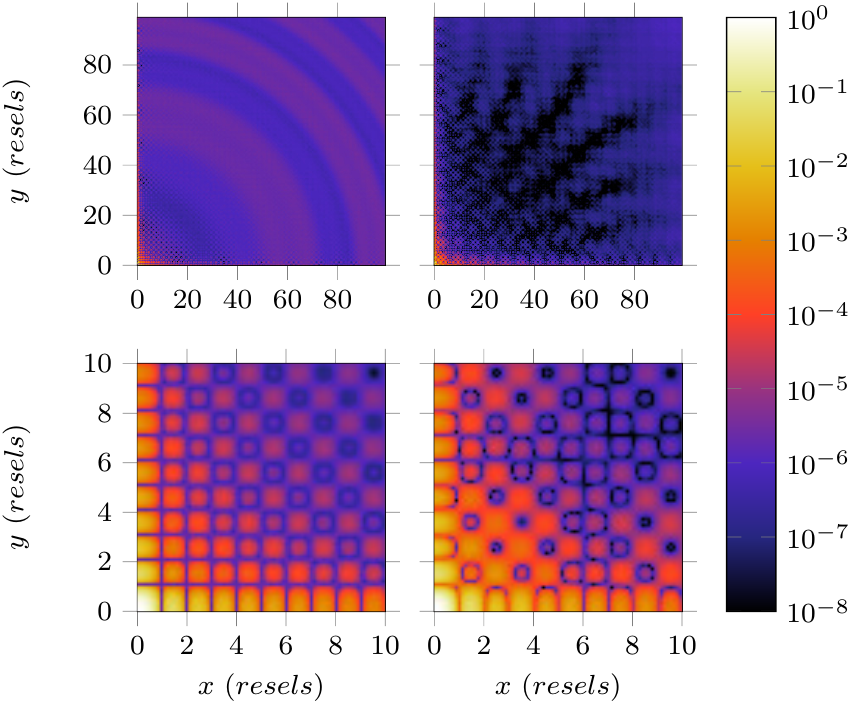}
	\caption{Quadrants of PSFs without central obturation (\textit{left}, $\epsilon=0$) and with central obturation (\textit{right}, $\epsilon=0.156$). 
		\textit{Top:} The 100 first resels from center. \textit{Bottom:} Zoom on the 10 first resels.
	}
	\label{fig:obt_psf}
\end{figure}

\begin{table}[htbp]
	\centering
	\caption{Performances of Fresnel arrays: with and without central obturation.}
	\label{table:obt_psf}
	\begin{tabular}{|l|||c|c|}
		\hline
		Obturation $\epsilon$ & $0$ & $0.156$\\
		\hline
		\hline
		\hline
		Raw transmission  $T$ & $49.94$ \% & $48.77$ \%\\
		\hline
		Efficiency $e$ at focal plane & $10.1$ \% & $9.84$ \% \\
		\hline
		Global dynamic range & $2.16\times10^1$ & $1.70\times10^1$\\
		\hline
		Dynamic range in $[5,15]$ resels & $5.93\times10^4$ & $8.83\times10^4$\\
		\hline
		Dynamic range in $[15,50]$ resels & $3.62\times10^5$ & $1.66\times10^6$\\
		\hline
	\end{tabular}
\end{table}

Table \ref{table:obt_psf} confirms that central obturation causes a loss of efficiency due to its blocking effect, however not critical. 

We see in Figure \ref{fig:obt_psf} that a central obturation covering $\epsilon=15.6~\%$ of the aperture has a positive impact on the PSF quality, as expected: it reduces the noise in the quadrants without significantly spreading the central lobe. The dynamic range is multiplied by 3 in the clean field between 15 and 50 resels.
We observe a sub-pattern with a period of $1/\epsilon = 6.41$ resels. An adapted apodization alleviates that. 

In the following, we use a central obturation with $\epsilon=0.156$.


\section{Holding bars}

For high contrast images we want the light to be focused only by diffraction: no optical material in the voids between Fresnel rings that could diffuse, alter the phase or affect the UV transmission. Still something has to hold the rings in place, such as a spider, the disadvantage of which being that it diffracts, thus pollutes the PSF. 

Ideally, pure rings could be held in place by a ``non-contact'' positioning: electromagnetic or radiation pressure. However, this would add technological burden for a space telescope that will already require formation flying if its focal length goes beyond 50 m, which is the case for large aperture Fresnel arrays.

For the moment we investigate on how to control and limit optically the unwanted diffraction caused by holding the rings, as we show in the following.

\subsection{Description of the rings holding system}

We test two different bars mesh: one with a pseudo-period, same as in the previous prototypes, and a new layout which is periodic. The PSF being altered by both, our goal is to reduce this alteration in the clean field.

\subsubsection{Pseudo-periodic bars layout}

Previous Fresnel arrays featured non-periodic orthogonal bars: tangent to their Fresnel rings, which are periodic as a function of their square radius. Bars and rings had a different pseudo-frequency:  one bar every $p_{bars}$ ring. We define -- $p_{bars}$ -- the pseudo-periodicity of bars as a function of rings. 
This previous pattern of bars corresponded to that of cylindric FZPs, but with unadapted focal lengths.

\begin{figure}[tbp]
   \begin{center}
   \begin{tabular}{ccc}
   \includegraphics[width=0.29\textwidth]{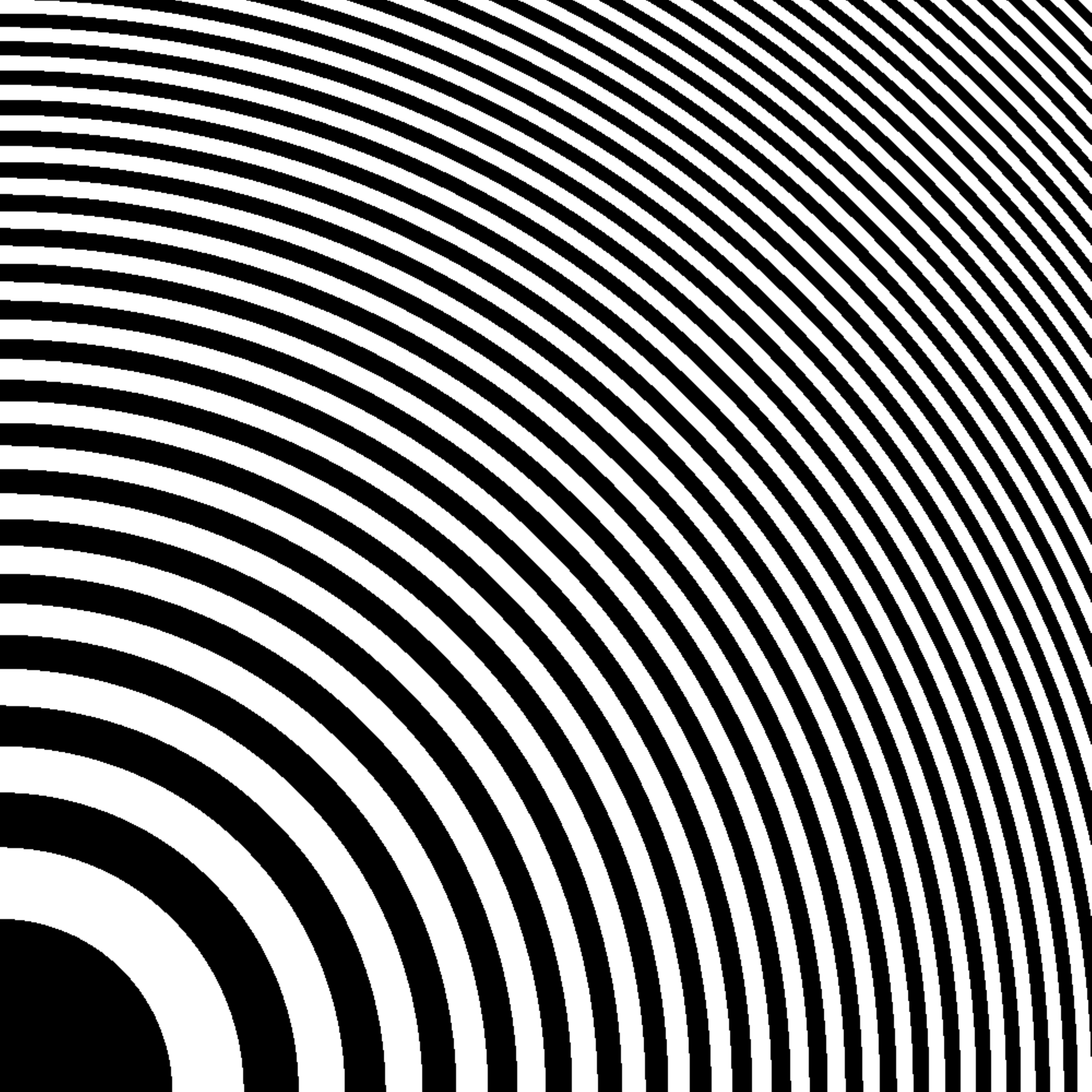} &
   \includegraphics[width=0.29\textwidth]{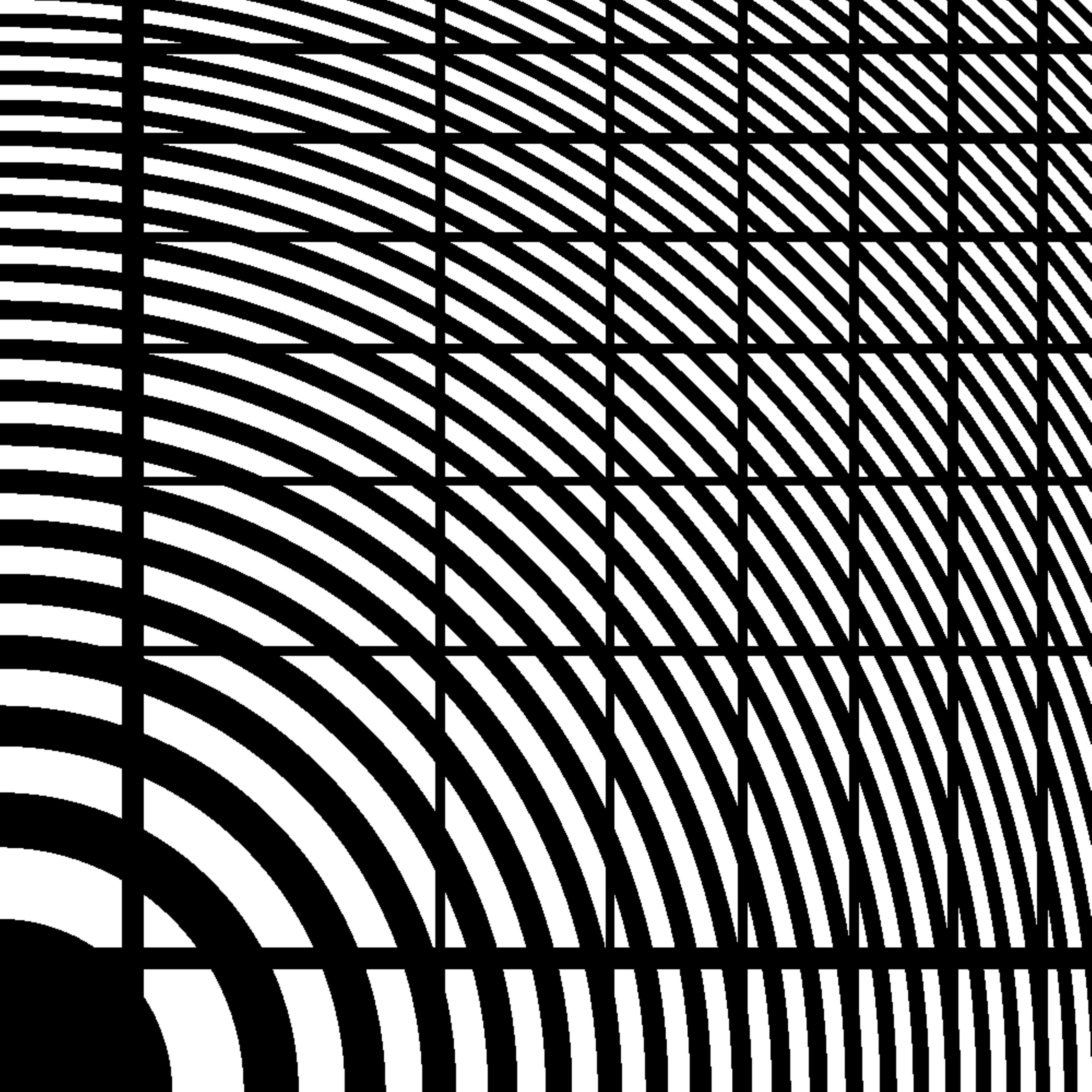} &
   \includegraphics[width=0.29\textwidth]{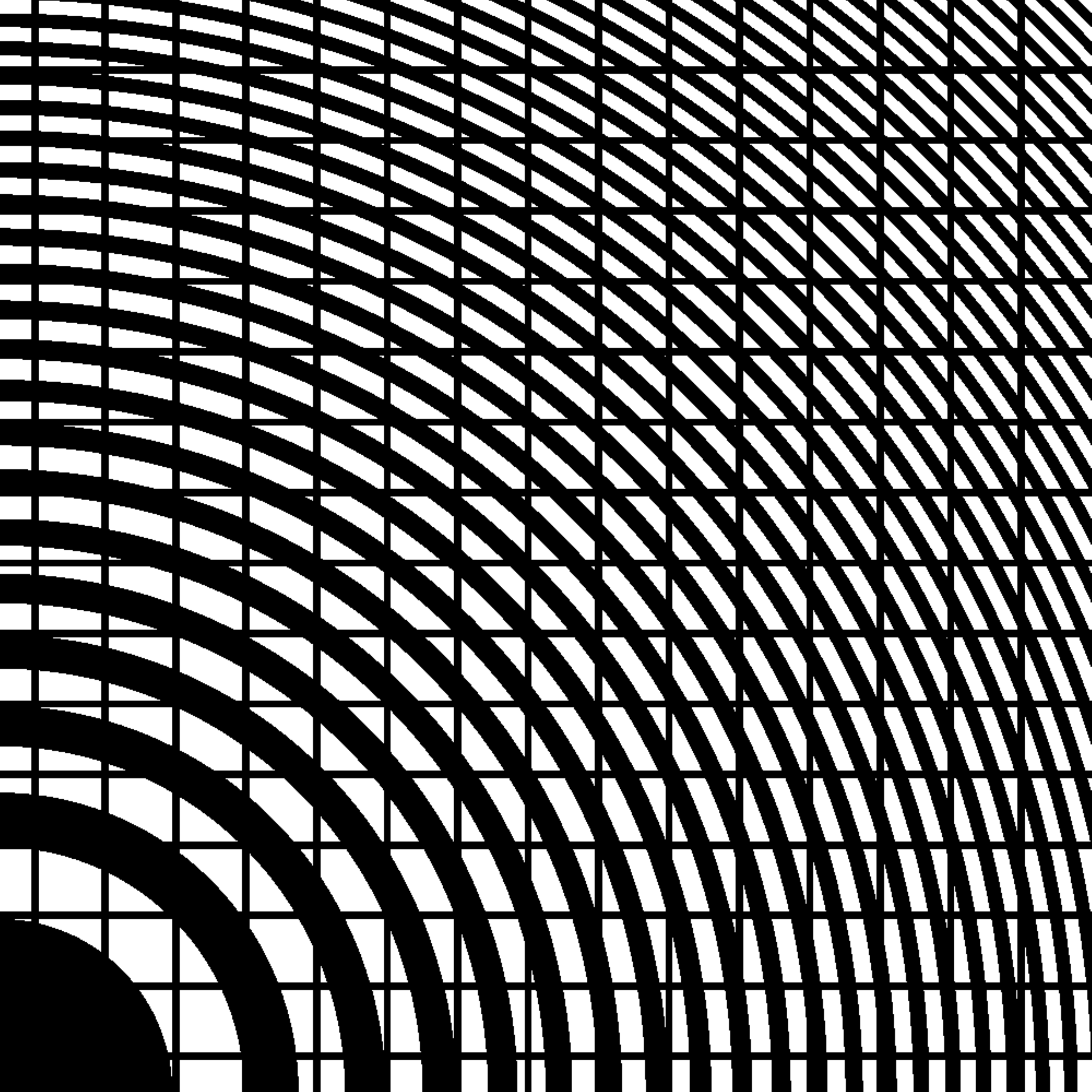} \\
   \end{tabular}
   \end{center}
   \caption[Examples of holding bars system.]{
   	Examples of holding bars for $N=20$ Fresnel rings from center to edge. We only show here the top right quadrant, the others are symmetric. \textit{Left:} No bars (transmission $T=49.55~\%$). \textit{Middle:} Pseudo-periodic bars with $p_{bars}=3$ ($T=42.81~\%$). \textit{Right:} Periodic bars with $N_{bars}=30$ ($T=40.12~\%$).
   }
   \label {fig:bars_ex} 
\end{figure}

One can  guess (and measure) that in addition to their effect on the PSF, bars reduce transmission, blocking a percentage of the light. To alleviate, we reduce the number of bars and modulate their widths (Fig.~\ref{fig:bars_ex}, \textit{middle}). 

The problem arising if we increase  $p_{bars}$ is mechanical instability, even though, due to the high focal ratio, that there is a large tolerance on the rings position for a given wavefront quality. Our previous optical tests have shown that we cannot use a pseudo-period superior to $p_{bars} = 3$ with ground based (Earth gravity) conditions. 
The minimum bars widths is limited by the stiffness of the plate into which the Fresnel array is cut. For ground-based conditions and $20\times20$ cm apertures, $d_{bars}$ has to be $\geq12~\mu m$ to avoid deformations that alters the wavefront (various plates 40 to $150~\mu m$ thick were tested in previous Fresnel arrays, in stainless steel, copper, CuBe alloy, polyimide, and copper-polyimide sandwich). 


\subsubsection{Periodic bars layout}

Now we compare the performances of a new bars design with the previous ones, using regularly spaced, narrow bars with $d_{bars}=12~\mu m$  (Fig.~\ref{fig:bars_ex}, \textit{right}).
 The bars-rings interaction casts a cleaner periodic dot pattern, into which the stray light is confined.
The period can be adjusted so that the resulting effect on the diffraction pattern is mostly rejected outside the clean field, yielding better images.

In a new design, equidistant bars in the aperture are defined by two parameters:
\begin{itemize}
    \item $d_{bars}$ their widths;
	\item $N_{bars}$ their number along one side.
\end{itemize}

Being a 2D grating, a bars mesh with a space periodicity of $p_{bars}$ will produce in the PSF an array of dots separated by:
\begin{equation}
\frac{\lambda f}{p_{bars}} = \frac{\lambda f (N_{bars}+1)}{d_{array}}
\end{equation}
One resel corresponding to a linear size of $\frac{\lambda f}{d_{array}}$, if we want no dots in a clean field $n$-resel wide from center we have to respect:
\begin{equation}
\frac{\lambda f (N_{bars}+1)}{d_{array}} > \frac{n \lambda f}{d_{array}} \Leftrightarrow N_{bars}+1 > n
\end{equation}
For instance, a clean field size of $100$ resels from center implies at least $N_{bars}=100$ per side.
In larger fields the diffraction pattern (dots) form the bars is present. However, it covers a very small fraction of the field and, same as for the spikes, it the clean field can be completed by a second exposure if necessary. 



\subsection{Results and discussion}

\begin{figure} [htbp]
	\centering
	\includegraphics{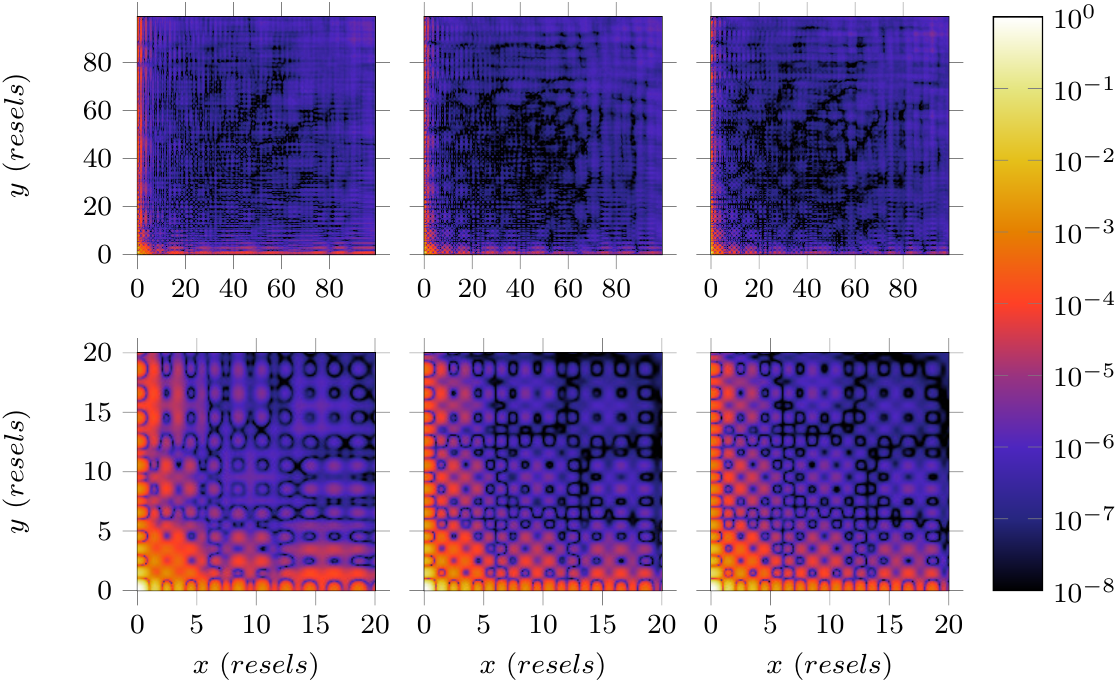}
	\caption{
		PSFs of Fresnel arrays with pseudo-periodic bars holding the rings.
		\textit{Left:} $p_{bars}=1$, \textit{Middle:} $p_{bars}=2$, \textit{Right:} $p_{bars}=3$. \textit{Top:} $100$-resel field, \textit{Bottom:} $10$-resel field near the central lobe.
	}
	\label {fig:per_bars_psf} 
\end{figure}

Pseudo-periodic bars holding the rings alter significantly the PSF:  in the first resels near center, adding noise (Fig.~\ref{fig:per_bars_psf}) 
and along the spikes. Fewer bars (higher values of $p_{bars}$) reduce these effects.


Table~\ref{table:results_per_bars} shows the luminosity efficiency and dynamic range as a function of $p_{bars}$ for pseudo-periodic bars. We are far from the theoretical efficiency of 10.1~\%, this is due to the combined effects of central obturation and bars width, although mostly bars, as with central obturation alone the efficiency reaches 9.84~\%.

The global dynamic range (relevant for extended surfaces or dense fields) is not reduced, it is even improved. But the high dynamic range capabilities in the clean field are reduced: respectively divided by 4 with $p_{bars}=1$ and by 1.7 with $p_{bars}=3$ (the latter setting is that of our previous, Mark II prototype).

\begin{table}[htbp]
	\centering
	\caption{Fresnel arrays performances with different pseudo-periodic bars holding the rings.}
	\label{table:results_per_bars}
	\begin{tabular}{|l|||c|c|c|c|}
		\hline
		Bars periodicity $p_{bars}$ & $1$ & $2$ & $3$ & $\infty$, no bars\\
		\hline\hline\hline
		Raw transmission $T$ & $31.41~\%$ & $39.42~\%$ & $42.32~\%$ & $48.77~\%$\\
		\hline
		Efficiency at focal plane  $e$ & $4.09~\%$ & $6.42~\%$ & $7.40~\%$ & $9.84~\%$\\
		\hline
		Global dynamic range & $2.70\times10^1$ & $2.04\times10^1$ & $1.88\times10^1$ &$1.70\times10^1$\\
		\hline
		Dynamic range in $[5,15]$ resels & $8.08\times10^4$ & $1.37\times10^5$ & $1.11\times10^5$ & $8.83\times10^4$\\
		\hline
		Dynamic range in $[15,50]$ resels & $4.48\times10^5$ & $1.09\times10^6$ & $9.49\times10^5$ & $1.66\times10^6$\\
		\hline
	\end{tabular}
\end{table}

\begin{figure} [htbp]
	\centering
	\includegraphics{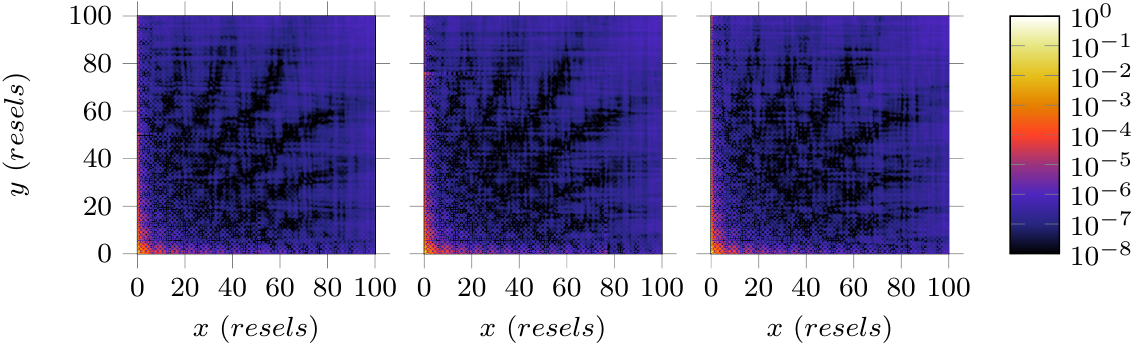}
	\caption{
	PSFs for truly periodic bars holding the rings: different bar frequencies (number of bars per side) compared at constant bar width $d_{bars} = 12~\mu m$:
		\textit{Left:} $N_{bars}=50$. \textit{Middle:} $N_{bars}=75$. \textit{Right:} $N_{bars}=100$.}
	\label {fig:reg_bars_psf} 
\end{figure}

\begin{table}[htbp]
	\begin{center}
		\caption{Fresnel array performances with truly periodic bars holding the rings, as a function of frequency $N$ (number of bars per side), at constant bar width $d_{bars} = 12~\mu m$.}
		\label{table:results_reg_bars}
		\begin{tabular}{|l|||c|c|c|c|}
			\hline
			$N_{bars}$ & $50$ & $75$ & $100$ & Without\\
			\hline\hline\hline
			Transmission rate $T$ & $47.87~\%$ & $47.47~\%$ & $47.00~\%$ & $48.77~\%$\\
			\hline
			Efficiency $e$ & $9.48~\%$ & $9.33~\%$ & $9.14~\%$ & $9.84\%$\\
			\hline
			Global dynamic range & $1.70\times10^1$ & $1.70\times10^1$ & $1.69\times10^1$ &$1.70\times10^1$\\
			\hline
			Dynamic range in $[5,15]$ resels & $9.00\times10^4$ & $8.46\times10^4$ & $7.75\times10^4$ & $8.83\times10^4$\\
			\hline
			Dynamic range in $[15,50]$ resels & $1.95\times10^6$ & $1.62\times10^6$ & $1.62\times10^6$ & $1.66\times10^6$\\
			\hline
		\end{tabular}
	\end{center}
\end{table}

~\

As a first step towards improvement, replacing pseudo-periodic by periodic bars, we see the following effects:
\begin{itemize}
\item {no notable change in the central $15$ resels.}
\item {The expected bright dots in the far field (Figure \ref{fig:reg_bars_psf}) in each ($N_{bars} + 1$)\textsuperscript{th} resels. }
\item {Except at these points the PSF background is smoother and the dynamic range almost twice better. }
 \item {The four spikes are well constrained, better than with pseudo-periodic bars.}
\end{itemize}
In Table~\ref{table:results_reg_bars}, we have set the bars width to ($d_{bars} = 12~\mu$m). Reducing the period of bars to reject further out their dots in the image slightly degrades light transmission, and has no effect on the PSF near the central lobe. The dynamic range is even better with $N_{bars}=50$ than without bars, but some bright dots 
remain. At $N_{bars}=100$, dots are chased out of the clean field, and on a different aspect the structure is mechanically more resistant. The performances slightly decrease compared to $N_{bars}=50$, but remain better than with the previous pseudo-periodic setup $p_{bars} = 3$.

So for the next step we keep a periodic bars mesh with the following parameters: $N_{bars}=100$ and $d_{bars} = 12~\mu$m, and work on other aspects to further improve the dynamic range.

\section{Apodization by transmittance modulation}

Apodization reduces the secondary lobes of the PSF, and so improves the dynamic range in the image. The way it is done usually reduces transmission, except for PIAA (Guyon et al. 2005) \cite{Guyon2005}. The crux of the matter is to get the best contrast with the best efficiency.
A classical apodization smoothly modulates the aperture transmittance $t(x,y)$ from center to edge. However, our Fresnel grids have a binary transmittance that can only be $0$ or $1$. 
We simulate a fractional transmittance by changing the relative size of voids between Fresnel rings (duty cycle): 
the voids in Fresnel zones are set to half the local pseudo-period where a maximum transmission is required, and they are thinned where the transmission to needs to be decreased. They could as well be enlarged  (G. Andersen, 2010)  \cite{Andersen2010} or locally displaced, which has an effect on the relative phase locally, thus on the integrated wave amplitude at focus. 
In the following we investigate the best apodization function for dynamic range in the images of Fresnel arrays.

\subsection{Our new  apodization}

The previous Fresnel arrays were already apodized but we now work on an orthogonal apodization, different from the radial law we used before.
Among different approaches (\textit{e.g.} PIAA (Guyon et al. 2005 \cite{Guyon2005}),  we have chosen the apodized square aperture (ASA), following Nisenson and Papaliolios (2001) \cite{Nisenson2001}, which is efficient enough for a small aperture and simple to implement.

\subsubsection{Apodization function}

\begin{figure} [!ht]
	\centering
	\includegraphics{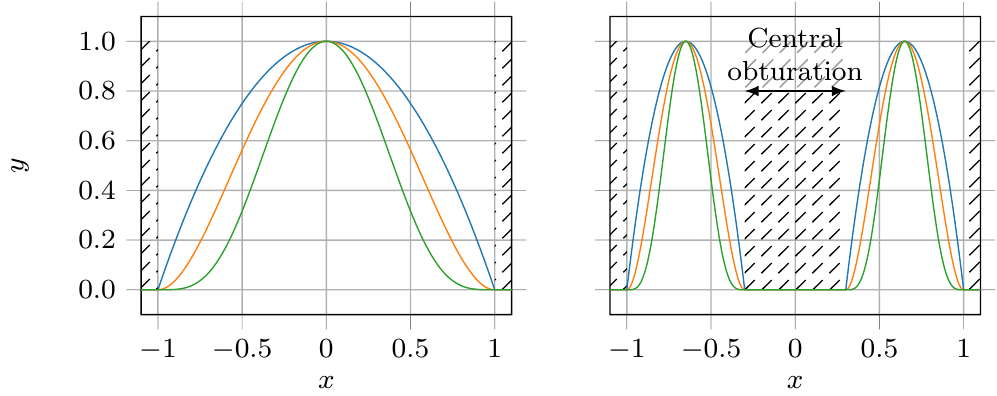}
	\caption[Sonine function and an adapted version for a central obturation.]{\textit{Left:} The sonine function as described by Nisenson and Papaliolios (2001) \cite{Nisenson2001} with three values of $\nu$ (\textit{Blue}: $\nu=2$, \textit{Orange}: $\nu=3$, \textit{Green}: $\nu=5$). \textit{Right:} Our adapted sonine function decreases to $f=0$ around the central obturation, to minimize its effect of the PSF (in this case $\epsilon=30\%$).
	}
	\label {fig:sonine} 
\end{figure}

The choice among different apodization functions for ASA depends on the desired dynamic range, transmission efficiency, and to some extent roughness of the PSF background. In our case we use a sonine function, defined for $x \in [-1, 1]$:
\begin{equation}
f(x) = (1 - x^2)^{\nu-1}
\label{eq:sonine} 
\end{equation}
$\nu > 1$ is a parameter defining the apodization strength (see Fig.~\ref{fig:sonine} left), $x$ is the position in the aperture : $x=0$ at center and $|x|=1$ at the edges.

We have a central obturation, so we adapted the sonine function. We start from a 1-dimensional apodization function (see Fig.~\ref{fig:sonine} right) as:
\begin{equation}
f_{obt}(x)=
\left\lbrace
\begin{array}{ccl}
\left[1- \left(\frac{2|x| - 1 - \epsilon}{1 - \epsilon}\right)^2\right]^{\nu-1}  &~& \mbox{if} ~ \epsilon < |x| \leq 1,\\
0 &~& \mbox{otherwise}.
\end{array}\right.
\end{equation}
$\epsilon$ is the central obturation (linear) ratio to the total aperture, so $\epsilon \in ]0,1]$.

\subsubsection{Two dimensional apodization}

\begin{figure} [tbp]
	\centering
	\includegraphics{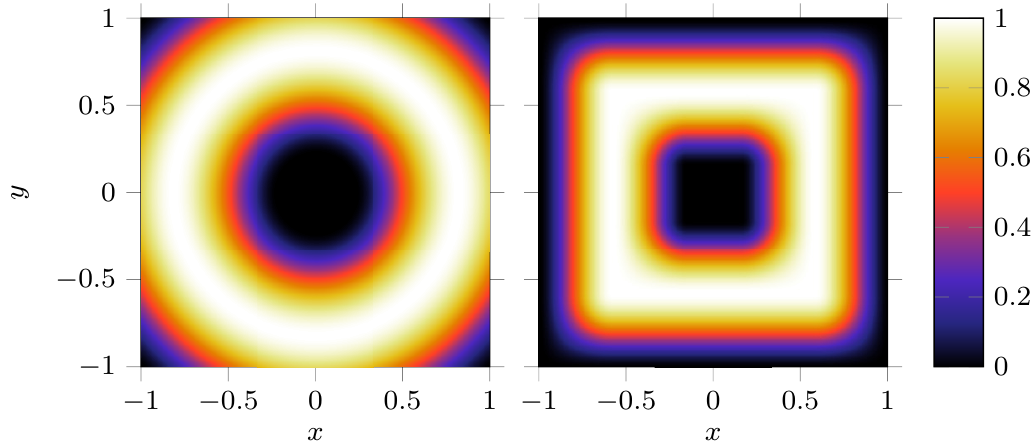}
	\caption{Two examples of apodized apertures with central obturation $\epsilon = 0.156$, and sonine function $\nu=3$. 
		\textit{Left:} Radial sonine apodization. \textit{Right:} Crossed sonine apodization.
	}
	\label {fig:apod2D} 
\end{figure}

\paragraph{Radial sonine apodization}
For a circular aperture where $(r, \theta)$ are the polar coordinates,
 the argument $r$ of the sonine apodization function varies from $-1$ to $1$ along a diameter:
\begin{equation}
f_{2D}(r, \theta) = f_{obt}(r) , ~ \forall ~ \theta
\end{equation}
If we use a circular sonine apodization  in a square aperture, $f_{obt}$ varying from $-1$ to $1$ along a diagonal, the transmission does not reach zero smoothly on the edges (see Fig.~\ref{fig:apod2D} left). However, the resulting PSF has some interesting characteristics, as shown later.

\paragraph{Crossed sonine apodization}
To better fit a square aperture, we can use a crossed sonine function as defined by Nisenson and Papaliolios (2001) \cite{Nisenson2001},  adapted for central obturation (see Fig.~\ref{fig:apod2D} right): 
\begin{equation}
f_{2D}(x, y) = \left\lbrace
\begin{array}{ccl}
1 - \left[\left(1-f_{obt}(x)\right)\left(1-f_{obt}(y)\right)\right] &~& 
\mbox{if}~|x| \leq \frac{1+\epsilon}{2} ~
\mbox{and}~ |y| \leq \frac{1+\epsilon}{2} ,\\
f_{ext}(x)f_{ext}(y) &~& \mbox{otherwise.}
\end{array}\right.
\end{equation}
where $f_{ext}$ is an extension of $f_{obt}$:
\begin{equation}
f_{ext}(x)=
\left\lbrace
\begin{array}{ccl}
1 &~& \mbox{if} ~ |x| < (1+\epsilon)/2,\\
\left[1- \left(\frac{2|x| - 1 - \epsilon}{1 - \epsilon}\right)^2\right]^{\nu-1}  &~& \mbox{if} ~ (1+\epsilon)/2 < |x| \leq 1,\\
0 &~& \mbox{otherwise}.
\end{array}\right.
\end{equation}

\subsubsection{Partial apodization}

Ideal apodization functions imply transmittances varying from $t=0$ to $t=1$. Because of a limit in manufacturing, a binary Fresnel array cannot reach $t=0$ just by modulating the duty ratio of the binary Fresnel rings: this would imply infinitely thin cuts, whereas the thinnest laser cutting tools available to us have a $\sim 20~\mu$m width. 


For our Fresnel array, the outer and narrowest half-zone measures $w_{320}= \frac{\sqrt{2}d_{array}}{16N} = 35.9~\mu m$, and considering a cutting width $w_{cutting} = 20~\mu m$. That's equivalent to a minimum local transmittance of $t_{min}=0.59$, assuming that the thinned half-zone remains centered on the original one.

A solution to reduce the light transmitted by the outer zones is  dashing them to reduce their transmission, 
as in the thin ``inter-petal'' ridges of stellar occulters from Cash (2011) \cite{Cash2011}.
We have not tested this method yet, but it is simple to implement.

\subsection{Results of apodization, and discussion}

In our apodized Fresnel arrays, we set the following: 
\begin{itemize}
\item {central obturation $\epsilon=0.156$;}
\item  {periodic holding bars mesh with $d_{bars} = 12~\mu m$;}
\item {$N_{bars} = 100$, for a 100-resel wide clean field.}
\end{itemize}

 \begin{figure} [htbp]
   \centering
   \includegraphics{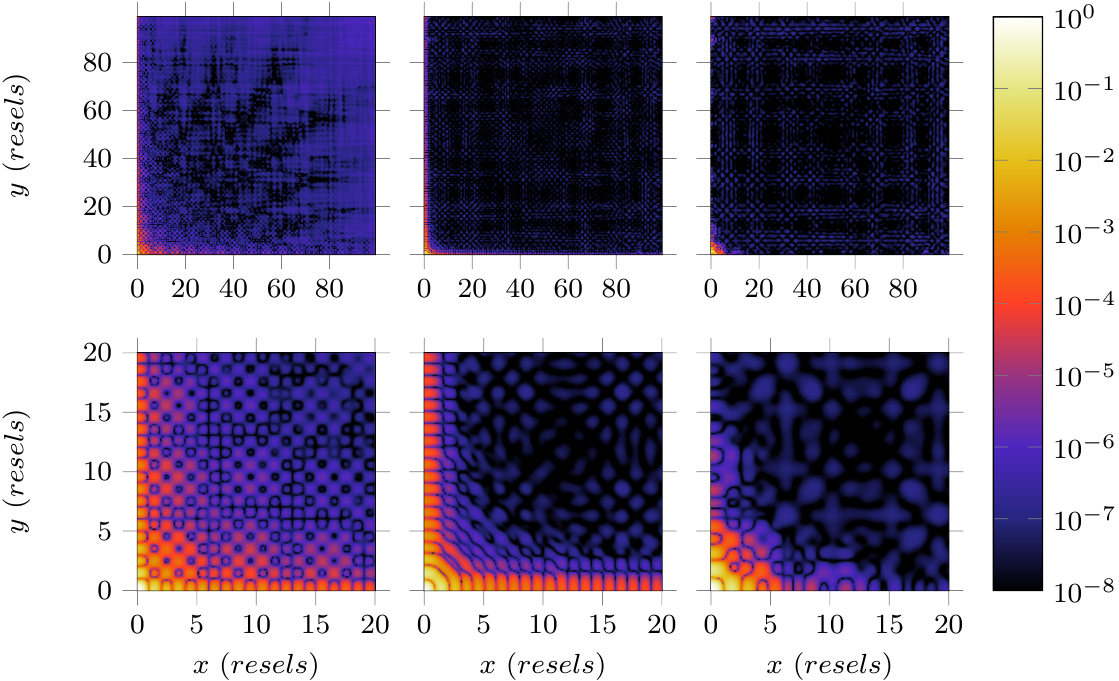}
   	\caption{PSFs of Fresnel arrays with different 2D sonine apodizations at $\nu = 3$ (see Equation \ref{eq:sonine}). \textit{Left:} No apodization, \textit{Middle}: Radial sonine,  \textit{Right:} Crossed sonine. \textit{Top:} $100$-resel field from the mail lobe, \textit{Bottom:} Zoom on the $20$ first resels from the main lobe. The PSF is symmetrical, so only the upper right quadrant is displayed.}
   \label{fig:apod_psf} 
\end{figure}

Fig.~\ref{fig:apod_psf} compares the PSF: without apodization and with 2 types of apodization. The angular resolution is preserved. 
Concerning the four spikes spreading from the central lobe, the radial sonine apodization leaves them high because it does not correct the aperture edges. On the contrary, with the crossed sonine apodization the spikes disappear after 15 resels from center. They are efficiently removed by a crossed sonine apodization, that has a smooth transition on the edges. As a consequence, square Sonine is adapted to imaging relatively extended objects.

\begin{table}[htbp]
\begin{center}
\caption{Transmission, efficiencies and dynamic ranges without and with 2D apodizations.}
\label{table:apod_psf}
\begin{tabular}{|l|||c|c|c|}
    \hline
    2D apodization function & Without & Radial sonine & Crossed sonine\\
    \hline \hline \hline
    Transmission through aperture $T$ & $47.00~\%$ & $32.49~\%$ & $25.08~\%$\\
    \hline
    Light efficiency at focus $e$ & $9.14~\%$ & $5.24~\%$ & $3.79~\%$\\
    \hline
    Global dynamic range & $1.69\times10^1$ & $8.88$ & $8.84$ \\
    \hline
    Dynamic range in $[5,15]$ resels & $7.75\times10^4$ & $3.74\times10^6$ & $4.45\times10^6$ \\
    \hline
    Dynamic range in $[15,50]$ resels & $1.62\times10^6$ & $4.98\times10^6$ & $5.34\times10^6$ \\
    \hline
\end{tabular}
\end{center}
\end{table}

Table~\ref{table:apod_psf} shows how raw transmission $T$ and efficiency $e$ at focus  are affected by the apodization. With apodization there is a much higher dynamic range in the close field (multiplied by $\approx 500$ between 5 and 15 resels), than without. Beyond 15 resels, the dynamic range continues to increase but slower. The quality of the far field is improved too (multiplied by $\approx 5$ between 15 and 50 resels). An adapted apodization yields theoretically to a very high dynamic range ($10^{15}$ in intensity) as shown by Nisenson and Papaliolios \cite{Nisenson2001}. But in our case, 
the perturbations from bars limit the dynamic range to the order of $10^7$ with a 160 Fresnel rings aperture. A larger number of Fresnel rings should improve that limit: our Mark II prototype has a 696-ring Fresnel array adapted to the visible, but its focal length is too long in UV for our tests.

\section{Optical tests on artificial sources}

The above work leads to the choice of a Fresnel array to manufacture for the next optical tests in UV. We vectorize this layout for production by laser machine tools: at present a Fresnel array adapted for UV has been carved, but not optimized by apodization. Manufacturing a Fresnel array is a matter of weeks in delivery time and a few hundred euros in cost. The other optical elements are ready to be tested. A new dispersion correction grating in reflexion is just finished, it will be used for the optical tests in the UV domain and will be the subject of an upcoming article.

In order to test the Fresnel imager in UV, we first prepare an observing run on artificial UV sources, where we will validate the PSF optically and confirm the numerical results. These tests will be carried in the next months, in the frame of our collaboration with the University of T\"{u}bingen, Germany, where we will use a photon counting UV camera based on a crossed stripes anode (CSA) and micro-channel plates stacks (MCPs) to detect the faintest stray light in the PSF, using high contrast test targets.

\section{Test mission proposed in space on the ISS, and Scientific return}
\label{s:iss}

\begin{figure} [htbp]
	\centering
	\includegraphics[scale=0.25]{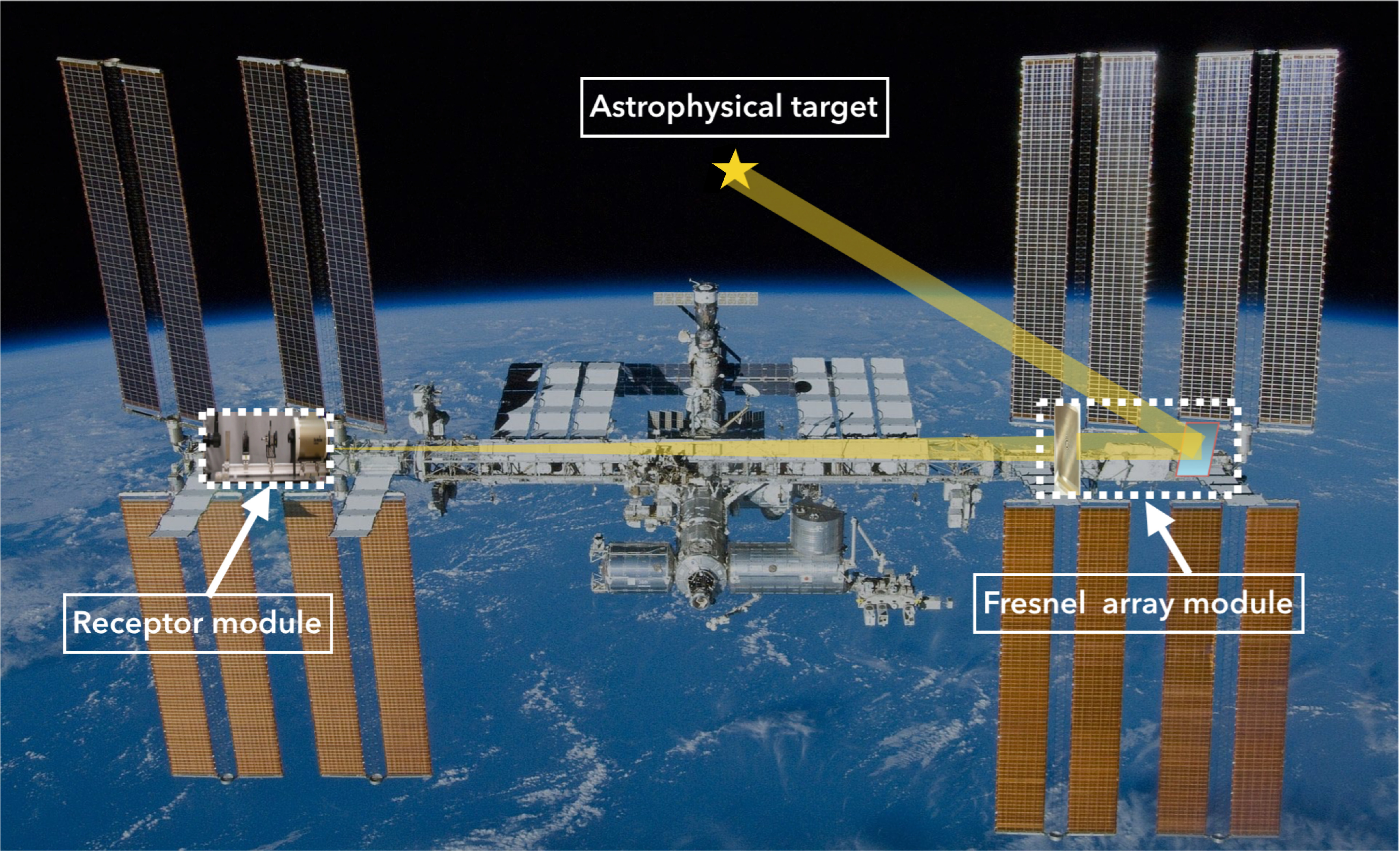}
	\caption{Sketch of the proposed probatory mission on the ISS (Credit for background image: NASA/Crew of STS-132). The focusing element: a Fresnel array, is in its module (on the \textit{right}). The field optics and the Fresnel mirror are in the receptor module (on the \textit{left}), the light beam is represented in yellow. A a siderostat mirror in front of the Fresnel array folds the beam, it allows target pointing and image stabilization for the tests on the ISS.}
	\label{fig:ISS} 
\end{figure}

Next we will apply for a test mission with a new Fresnel imager prototype (Mark IV), that we will propose to place on the ISS for a test run on astrophysical sources in UV. Specifications are detailed in Table~\ref{table:specs_prob_mission}.

\begin{table}[htbp]
	\centering
	\caption{\label{table:specs_prob_mission}Specifications of the future proposed probatory mission.}
	\begin{tabular}{|l|c|}
		\hline
		\multirow{2}{*}{Wavelength range} & 120-200 nm \\ & (includes Lyman-$\alpha$ at $\lambda$ =121 nm)\\
		\hline
		Aperture  $d_{array}$& 15 cm \\
		\hline
		number of Fresnel zones  $N$ & 500\\
		\hline
		Distance: Fresnel lens to field optics & 45 m \\
		\hline
	\end{tabular}
\end{table}

The probatory mission that we propose consists of 2 modules, as shown in the Figure \ref{fig:ISS}. The Fresnel array module contains the primary aperture, fixed. That's why it also contains a siderostat to aim at different targets and compensate for the drifts and vibrations of the ISS. The second module contains the field optics, chromatic corrector and focal instruments : a UV camera and  a spectro-imager for example).

We had similar action plans for our previous validations in the visible: we tested our Fresnel imager (Mark II) on astrophysical targets (Sirius AB, Mars satellites, M42 and $\theta$ Ori). This has been much more informative than the previous tests on artificial sources. We plan to do the same for UV with various astrophysical targets. These tests have three aspects:
 \begin{itemize} 
\item {angular resolution;}
\item {low contrast, (relatively) large field imaging;}
\item {high contrast imaging at narrow field.}
\end{itemize}

For each of these we will choose adapted astrophysical targets, and if possible bring along some scientific returns: even though we plan to have a modest $15$ cm aperture, the high contrast capabilities associated with a 0.15 arc second resolution in Lyman-$\alpha$ should allow some new results for well-chosen science cases such as close stellar environments and solar system objects (Koechlin and Berdeu 2014 \cite{Koechlin2014b}).

Once validated in relevant space conditions, we plan to apply for a full sized UV imager that could be used as space telescope for a broad set of astrophysical science cases requiring high dynamic range in UV (see Raksasataya et al. 2011 \cite{Raksasataya2011}, G{\'o}mez de Castro 2011 \cite{GomezdeCastro2011}, Hinglais 2011 \cite{Hinglais2011}).

\section{Conclusion}

From this study on our ground prototype, we see that for holding the Fresnel rings, a regular spacing of bars provides better images than the pseudo-periodic spacing in our previous prototypes. The crossed sonine apodization is clearly better for extended sources, and provides a higher dynamic range for narrow fields despite the efficiency loss in luminosity. To keep the high quality wavefront allowed by binary transmission (opaque an void rings), we will apply the apodization by modulation of the rings duty ratio, as we already had done in our previous prototypes.

Thanks to this numerical propagation tool adapted for imaging by Fresnel arrays, the design is optimized before manufacturing. The optical tests are planned in Germany in the next few months. 
Our prototype is tested at $\lambda = 260$ nm, although we plan to propose science cases as far in UV as Lyman-$\alpha$ for the large Fresnel arrays, as well as for the probatory mission on the ISS: the present work helps to validate a new optical design and chromatic correction for UV, which will remain similar at shorter wavelengths.
%


\bibliographystyle{plain}
\bibliography{main} 

\end{document}